\documentclass[preprint2]{aastex}

\newcommand{\hh}    {H$_2$}
\newcommand{\brgamma}   {Br$\gamma$}
\newcommand{\vone}  {\hbox{1--0 $S(1)$}}
\newcommand{\vonez} {\hbox{1--0 $S(0)$}}

\newcommand{\vtwo}  {\hbox{2--1 $S(1)$}}
\newcommand{\vthree}{\hbox{3--2 $S(3)$}}
\newcommand{\vfive} {\hbox{5--3 $O(3)$}}

\newcommand{\jone}  {$J = 1 \rightarrow 0$}
\newcommand{\jtwo}  {$J = 2 \rightarrow 1$}
\newcommand{\jthree}{$J = 3 \rightarrow 2$}
\newcommand{\jfour} {$J = 4 \rightarrow 3$}

\newcommand{\hi}{\hbox{H\,{\scriptsize I}}}
\newcommand{\hii}{\hbox{H\,{\scriptsize II}}}

\newcommand{\oi}{\hbox{[O\,{\scriptsize I}}]}
\newcommand{\cii}{\hbox{[C\,{\scriptsize II}}]}

\newcommand{\kms}{km~s$^{-1}$}
\newcommand{\wattmeter}{W~m$^{-2}$}

\newcommand{\lsun}{$L_{\odot}$}

\newcommand{\av}{$A_{V}$}

\slugcomment{to appear in Astrophysical Journal}

\shorttitle{H$_2$ emission from external galaxies}
\shortauthors{Pak, Jaffe, and Stacey et al.}

\begin{document}


\title{
    Near-Infrared Molecular Hydrogen Emission from  \\
    the Central Regions of Galaxies: \\
    Regulated Physical Conditions in the Interstellar Medium
}

\author{
    Soojong Pak\altaffilmark{1,2,3},
    D. T. Jaffe\altaffilmark{3,4},
    G. J. Stacey\altaffilmark{5}, \\
    C. M. Bradford\altaffilmark{5,6},
    Eric W. Klumpe\altaffilmark{4,7},
    and
    Luke D. Keller\altaffilmark{3,4,8}  \\
  . \\
   {\it to appear in Astrophysical Journal}
}


\altaffiltext{1}{
     Korea Astronomy Observatory,
     61-1 Whaam-Dong, Yuseong-Gu, Daejeon 305-348, South Korea,
     $soojong@kao.re.kr$ }
\altaffiltext{2}{ Visiting Assistant Professor,
    The Institute of Space and Astronautical Science,
    Kanagawa 229-8510, Japan }
\altaffiltext{3}{ Visiting Astronomer,
    Cerro Tololo Inter-American Observatory and
    Kitt Peak National Observatory,
    National Optical Astronomy Observatory, which is operated by
    the Association of Universities for Research in Astronomy, Inc. (AURA)
    under cooperative agreement with the National Science Foundation. }
\altaffiltext{4}{ Department of Astronomy,
    The University of Texas, Austin, TX 78712 }
\altaffiltext{5}{ Department of Astronomy, Cornell University,
    230 Space Sciences Bldg., Ithaca, NY 14853 }
\altaffiltext{6}{ Jet Propulsion Laboratory,
    California Institute of Technology Pasadena, CA 91109 }
\altaffiltext{7}{ Department of Physics and Astronomy,
    Middle Tennessee State University, Murfreesboro, TN 37132 }
\altaffiltext{8}{ Department of Physics, Ithaca College, CNS 264,
Ithaca, NY 14850 }

\begin{abstract}
The central regions of many interacting and early-type spiral
galaxies are actively forming stars. This process affects the
physical and chemical properties of the local interstellar medium
as well as the evolution of the galaxies. We observed
near-infrared \hh\ emission lines: $v=$ \vone, \vthree, \vonez,
and \vtwo\ from the central $\sim 1$~kpc regions of the
archetypical starburst galaxies, M82 and NGC 253, and the less
dramatic but still vigorously star-forming galaxies, NGC 6946 and
IC 342. Like the far-infrared continuum luminosity, the
near-infrared \hh\ emission luminosity can directly trace the
amount of star formation activity because the \hh\ emission lines
arise from the interaction between hot and young stars and nearby
neutral clouds. The observed \hh\ line ratios show that both
thermal and non-thermal excittion are responsible for the
emission lines, but that the great majority of the near-infrared
\hh\ line emission in these galaxies arises from energy states
excited by ultraviolet fluorescence. The derived physical
conditions, e.g., far-ultraviolet radiation field and gas density,
from \cii\ and \oi\ lines and far-infrared continuum observations
when used as inputs to photodissociation models, also explain the
luminosity of the observed \hh\ \vone\ line. The ratio of the \hh\
\vone\ line to far-IR continuum luminosity is remarkably constant
over a broad range of galaxy luminosities; $L_{H2}/L_{FIR}$
$\simeq 10^{-5}$, in normal late-type galaxies (including the
Galactic center), in nearby starburst galaxies, and in luminous IR
galaxies (LIRGs: $L_{FIR}$ $>$ $10^{11}$~\lsun). Examining this
constant ratio in the context of photodissociation region models,
we conclude that it implies that the strength of the incident UV
field on typical molecular clouds follows the gas density at the
cloud surface.
\end{abstract}


\section{INTRODUCTION}

The {\it Infrared Astronomical Satellite (IRAS)} detected a
significant population of galaxies that emit large fractions of
their bolometric energy in the mid- and far-infrared (IR) bands.
The majority of luminous IR galaxies (LIRGs: $L_{FIR}$ $>$
$10^{11}$~\lsun ) are powered by massive stars that have recently
formed in their central kiloparsec \citep{Genzel98}. The far-IR
luminosities of these galaxies therefore measure the amount of
star formation activity \citep{Kennicutt98}. Local starburst
galaxies with far-IR luminosities of $10^{10} - 10^{11}$ \lsun \
and normal spiral galaxies with $L_{FIR}$ $\simeq 10^9$ \lsun \
also show evidence for active star formation in their central ($<
1$~kpc) regions, even though their far-IR luminosities are much
lower than those of LIRGs \citep{Thronson87,Stacey91}.

The formation of O and B stars affects the physical and chemical
state of the interstellar medium through stellar UV radiation and
through the effects of stellar winds and supernovae
\citep{Telesco88}. Strong near-IR \hh\ emission lines have been
observed from the central regions of galaxies, especially from
starburst galaxies and LIRGs. Since \hh\ vibration-rotation
emission can trace the interaction of the neutral ISM with the UV
radiation from hot young stars and its interaction with sources of
dynamical energy such as supernovae, the \hh\ luminosity may also
be a good tracer of star formation activity. The main motivation
of this paper is to explore how well we can use this additional,
readily observable tracer of the interaction of massive stars with
the dense interstellar medium to study massive star formation in
the inner parts of galactic disks.

The observational part of this paper emphasizes new, large-beam
observations of emission in the vibration-rotation transitions of
\hh\ in the near-IR from nearby late-type galaxies and low
luminosity starbursts. For normal galaxies, observations of
near-IR \hh\ lines have been restricted to local bright regions
because the average surface brightness of the \hh\ emission is
low. In a previous study, we used an instrument optimized for
observations of low surface brightness line emission in the
near-IR to detect enormously extended ($\sim 3\degr$ or $\sim
400$~pc) \hh\ emission along the plane of the Galaxy near the
Galactic center \citep{Pak96a,Pak96b}. The integrated and
dereddened \hh\ $v=$ \vone\ luminosity from the extended region
($8 \times 10^3$ \lsun) dominates over that from the previously
known 5~pc circumnuclear gas ring ($4.5 \times 10^2$ \lsun,
\citealt{Gatley84}).

Our results for the inner 400 pc of the Milky Way motivated us to
begin a study of \hh\ emission from the central kiloparsec of
other nearby galaxies where it may be possible to detect extended,
low surface brightness \hh\ emission using the same large-aperture
Fabry-Perot spectrometer used in the Galactic work. The large-beam
observations we present here make it possible to compare for the
first time the \hh \ emission to the far-IR continuum and far-IR
fine-structure line emission at similar angular resolution and
over similar angular scales. The observations also give us
measurements of \hh \ line strengths from nearby objects with
linear scales comparable to those obtained with slit spectrometers
for the more distant luminous IR galaxies. With this broad range
of spatially compatible observations, we investigate whether it is
possible to formulate a generalized star formation law which can
be applied to galaxies over a wide range of levels in star
formation activity.

In order to analyze the \hh\ line data and to compare them with
other star formation tracers, we need to identify the  excitation
mechanism responsible for the \hh\ emission. In the central
regions of galaxies, however,  the  excitation mechanism for the
\hh \ emission has long been the subject of intense controversy
(see the reviews of \citealt{Mouri94}; \citealt{Goldader97a}).
\citet{Puxley88} observed nearby spiral galaxies using a circular
variable filter ($\lambda/\Delta \lambda = 120$) and a beam
full-width at half maximum (FWHM) of 19\farcs 6. They detected
\vone, \vtwo, and \vonez\ lines and concluded that the \hh\ line
populations are non-thermal. However, \citet{Moorwood90} presented
high spectral resolution data ($\lambda/\Delta \lambda = 1500$)
and pointed out that the \vtwo\ flux measurement of
\citet{Puxley88} could be affected by unresolved stellar
absorption lines. They argued that the \hh\ emission is mostly
from shocks. \citet{Mouri94} suggested that high spectral
resolution observations are essential to discriminate between the
excitation mechanisms.

The ultimate energy source responsible for the excitation of the
\hh\ vibration-rotation emission can lie either in the absorption
of UV photons from the ambient interstellar radiation field or
from local OB stars, or in shock waves excited by supernovae,
protostellar flows, or other energetic events
\citep{Black87,Burton92}. Hydrogen molecules that absorb far-UV
radiation make a transition to an excited electronic state.  At
low densities, this transition is followed by fluorescence to the
ground electronic state and a cascade through that state's
vibration-rotation levels \citep{Sternberg89a,Pak03}. This
interaction between \hh\ molecules and UV radiation takes place at
the edges of molecular clouds where hydrogen makes the transition
from predominantly atomic to predominantly molecular gas, in the
photodissociation region (PDR: \citealt{Tielens85};
\citealt{Hollenbach99}). If $n_H$ $<$ $5 \times 10^4$~cm$^{-3}$,
the relative intensities of vibration-rotation transitions in the
near-IR arising in UV-excited \hh\ are insensitive to density or
to UV field strength \citep{Black87} and depend only on the
branching ratios in the downward cascade. In dense ($n_H$ $>$ $5
\times 10^4$~cm$^{-3}$) gas heated by intense far-UV radiation,
\citet{Sternberg89b} showed that inelastic H$-$\hh\ and \hh$-$\hh\
collisions can heat the gas at the cloud edge to temperatures
approaching 1000~K. A combination of the collisional de-excitation
and collisional excitation in the high temperature layer alters
the \hh\ level populations in this UV-excited gas. The expected
changes in relative line strength with density have been verified
by observations of molecular cloud edges \citep{Luhman97}. In
shocked regions, the \hh\ vibration-rotation energy level
populations in the ground electronic state are thermalized by
collisions and the level populations are usually in LTE
\citep{Shull82,Black87}. Because the level populations are
thermalized and reflect a similar range of temperatures,
shock-heated gas and hot, dense gas heated by UV radiation are
difficult to distinguish.

Most previous observations of the central regions of galaxies
showed that the \hh\ line ratios of \vtwo\ to \vone\ lie between
the ratios one would expect for fluorescently excited \hh\ and
what one would see in observations of UV-heated but collisionally
excited gas or of gas collisionally heated in shocks
\citep{Doyon94,Kawara90,Koornneef96,Lumsden94,Moorwood90,Puxley88,Puxley90,Davies03}.
These intermediate ratios suggest that both pure UV-fluorescence
and thermalized UV-excited (or shock excited) gas are responsible
for the \hh\ emission \citep{Davies03}. Detection of \hh\ lines
arising from upper vibrational states with $v > 2$ is a strong
indication of the presence of UV-fluorescent excitation since the
high energies of the upper states make it very hard to excite
these transitions thermally. There are marginal detections of such
transitions from a few galaxies, e.g., a giant \hii\ region NGC
5451 in M101 \citep{Puxley00}, Arp 299 \citep{Sugai99}, and some
ultraluminous IR galaxies \citep{Davies03}. Another method for
discriminating between the \hh\ excitation mechanisms is to
measure the \hh\ nuclear spin degeneracy ratio (ortho- to
para-ratio), which may deviate from its formation value if the gas
is thermally processed, but remains constant in the case of
non-thermal excitation. Observations of \vone\ and \vonez\ showed
that the \hh\ emission in the galactic centers arises mostly from
UV-fluorescence \citep{Harrison98,Sugai99,Puxley00}. More
recently, \citet{Davies03} have observed vibration-rotation
transition lines of \hh\ from the nuclei of seven ULIRGs and
concluded that the line emission arises from dense PDRs irradiated
by intense far-UV radiation fields. In galactic sources, we have
used observations of \hh\ lines from high vibrational states
\citep{Luhman95}, as well as comparisons of extended \hh\ emission
and tracers of UV excited PDRs such as far-IR continuum and \cii\
emission to show that much of the \hh\ emission arises from
UV-excited gas \citep{Luhman96,Luhman97,Klumpe99}. This comparison
with far-IR lines and continuum has been more difficult in other
galaxies because of the significant differences between the \cii\
and far-IR\ beam sizes and the typical slit sizes used for near-IR
spectroscopy of \hh. In this work, we will use both \hh\ line
ratios and a study of various tracers of star forming clouds in
the context of PDR models to learn more about the relative
significance of various types of \hh\ excitation.

In Section~\ref{sec:Observations}, we present a description of our
spectroscopic observations of the starburst galaxies M82 and NGC
253, and the normal spiral galaxies NGC 6946 and IC 342. In
Section~\ref{sec:Results} we first show the observed \hh\ spectra
and their extinction-corrected line fluxes. We then show the
spatial distribution of the \hh\ \vone\ line intensity and compare
it to the distribution of other PDR tracers, e.g., CO line
emission and sub-mm continuum. In Section~\ref{sec:Discussion} we
analyze the nature of the \hh\ excitation mechanism by comparing
the measured \hh\ line ratios with thermal and non-thermal models.
We also calculate the luminosity ratios of \hh\ to far-IR and
\cii\ to far-IR, and discuss the implications of the observed
ratios by comparison to predictions by PDR models.
Section~\ref{sec:Conclusions} summarizes our conclusions.

\section{OBSERVATIONS} \label{sec:Observations}

\subsection{Fabry-Perot Spectrometer}

We have observed emission lines resulting from vibration-rotation
transitions of \hh\ from the central $0.3-1.5$ kpc regions of M82,
NGC 253, NGC 6946, and IC 342. Table~\ref{tbl-1} lists the basic
parameters of the four galaxies in our sample. We made the
observations at the 1.5 m telescope of the Cerro Tololo
Inter-American Observatory (CTIO) and the 5 m Hale telescope of
the Palomar Observatory\footnote{
    Observations at Palomar Observatory were made as part of a
    continuing collaborative agreement between the California
    Institute of Technology, Cornell University, and the Jet
    Propulsion Laboratory}
with the University of Texas Near-IR Fabry-Perot Spectrometer
(UTFPS). The spectrometer was designed to observe extended and low
surface brightness line emission by maximizing the area-solid
angle product \citep{Luhman95}. In order to subtract the telluric
OH line emission, we chopped the secondary mirror from the source
position to $\Delta \alpha = \pm 16\arcmin$ (at CTIO) or $\pm
2\arcmin$ (at Palomar) at 0.5 Hz. The Fabry-Perot etalon was
scanned to cover the emission profiles, and each spectral step
consisted of one chopping cycle: object--sky--sky--object (see
Table~\ref{tbl-2}). The detailed procedure for the UTFPS
observations is described in \citet{Pak98}. The instrument, as
used for the observations described in this paper, differed in one
respect from the instrument described in \citet{Luhman95}. For the
1996 September run, the Fabry-Perot etalon was housed in an
enclosure cooled thermoelectrically to $-10 \degr$C to reduce the
background signal and to stabilize the instrument against drifts
in the plate spacing.

The beam shape on all runs was approximately cylindrical
\citep{Pak98}. We indicate the beam  size at each telescope using
the equivalent disk size, $\theta_{ED}$, the diameter of a
cylindrical beam whose solid angle is the same as the integrated
solid angle of the actual beam pattern. Since the one-dimensional
beam pattern was close to that of a box function, the FWHM of the
beam pattern is almost the same as $\theta_{ED}$. After the 1995
December run at the Palomar Observatory, we changed the optical
configuration inside the instrument. The equivalent disk size on
the 1996 September run ($\theta_{ED} = 11\arcsec$)  was smaller
than that on the 1995 December run ($\theta_{ED} = 16\arcsec$).
The telescope pointing uncertainty was, in all cases, less than 10
percent of the beam size. As the area-solid angle product of the
instrument changes, the resolving power of the spectrometer also
changes by a small amount, i.e., R=$\lambda/\Delta \lambda$ varies
between 2300 and 3000. Table~\ref{tbl-2} therefore also lists the
effective spectral resolution for each wavelength and observing
run. The flux calibration was done by measuring standard stars at
the center of the beam profile. The flux calibration uncertainty
is less than 20 percent. Table~\ref{tbl-2} summarizes the
observation logs, and lists the instrument parameters and the flux
calibration stars.

On the 1995 December run and the 1996 September run, we used the
Hale telescope, where the UTFPS beam has a linear size of
$100-300$~pc in our target galaxies, to map the central regions of
these galaxies (except NGC 6946, where only a single point was
observed). We observed \hh\ \vone\ line emission along the major
axis or the molecular bar known previously from CO observations
(see Table~\ref{tbl-1}) stepping either 10\arcsec \ or 15\arcsec\
between the individual measurements. At some positions, we also
observed \vtwo\ line emission. On the 1995 October CTIO run, we
observed the \vone, \vtwo, and \vthree\ lines in NGC~253 with a
beam size of 81\arcsec\ ($\sim$1 kpc in this galaxy). This large
beam covers the entire region scanned during the 11\arcsec\ beam
observations in NGC~253.

\subsection{Long-slit Spectrometer}

We observed \hh\ emission lines from M82 with the IR Cryogenic
Spectrometer (CRSP; \citealt{Joyce94}) at the 2.1~m telescope of
the Kitt Peak National Observatory (KPNO) in 1996 December. The
CRSP is a long-slit spectrometer with a slit length of 82\farcs4
(135~pixels $\times$ 0\farcs61). We used a 300 l/mm grating and a
slit width of 2.7 pixels (or 1\farcs65) and the resulting spectral
resolution was $\Delta V_{FWHM}$ $=$ $283$~\kms\ at $2.2 \micron$,
corresponding to R$\simeq$1060. The grating was tilted to take
spectra in the $2.11-2.25$~\micron\ band which includes the \hh\
\vone\ 2.12183 \micron, \vonez\ 2.22329 \micron, and \vtwo\
2.24771 \micron\ emission lines.

The CRSP slit used toward M82 was rotated to match the major axis
(P.A. = 73\degr) of the CO molecular bar \citep{Shen95}. In order
to cover the extended emission from the central region, we also
took spectra with the slit displaced by $\pm 2~\arcsec$
perpendicular to its length: ($\Delta \alpha$, $\Delta \delta$)
$=$ ($-0\farcs58$, $-1\farcs91$), (0, 0), and ($+0\farcs58$,
$+1\farcs91$). We co-added 9 pixels along the spatial direction
(or slit length direction) and co-added the data taken at all
three slit positions to make high signal-to-noise spectra
reflecting the near-IR emission from a rectangular beam of
$5\farcs6 \times 5\farcs5$. In the central region of M82, the \hh\
emission may extend up to 10\arcsec\ in the minor axis direction
(see Section~\ref{sec:Distr_M82}). Even though the co-added
rectangular beam did not include all of the flux from M82, the
\hh\ line ratios from these data are more reliable than those
derived from Fabry-Perot observations made at different times
because the ratios are less dependent on the sky conditions (see
Section~\ref{sec:H2Ratio}). The CRSP measurements can therefore
contribute to our analysis of the \hh\ excitation mechanism.

\section{RESULTS} \label{sec:Results}

\subsection{ \hh\ Spectra} \label{sec:H2Spectra}

Figures $\ref{fig-01} - \ref{fig-07}$ show the \hh\
spectra taken with the UTFPS toward M82, NGC 253, NGC 6946, and IC 342.
On each \hh\ spectrum, we have overlaid a $^{12}$CO \jone\
spectrum taken with a similar beam size and smoothed in the
spectral direction with a Gaussian smoothing function to the
resolution of the UTFPS to allow us to compare the radial
velocities and the line shapes more carefully. In general,
those \hh\ line profiles with high signal-to-noise are
similar to the CO profiles from the same positions. The similarity
argues for a similar distribution of line emissivity with
galactocentric radius, implying that the \hh\ emission and CO
emission arise from the same molecular clouds. At the M82
positions (E27\arcsec,N13\arcsec) and (W27\arcsec, S13\arcsec),
however, the CO and \hh\ profiles differ.  At least part of the
change in shape and velocity profile may result from
the beam pattern difference between the CO and \hh\ observations.
The spatial 16\arcsec\ beam of the Nobeyama Radio
Telescope is Gaussian while that of the UTFPS 16\arcsec\ beam is
cylindrical. This shape difference could work in concert with the large
gradient in rotational velocity at these positions to produce
the profile differences.

Figure~\ref{fig-08} shows selected spectra taken by CRSP toward
M82. The \hh\ \vone\ lines are prominent and detected at all
positions. The \vtwo\ and \vonez\ lines are clearly detected at
the (E10\farcs5, N3\farcs2), (W10\farcs5, N3\farcs2), and
(W21\farcs0, S6\farcs4) positions, but are confused with stellar
absorption features at the (0, 0) position where the continuum
level is highest. The relative strength of \vonez\ and \vtwo\ in
this work agrees with that of \citet{Foerster01} (see
section~\ref{sec:H2Ratio}). There is a possible detection of the
\vthree\ line at (W21\farcs0, S6\farcs4), but this line was not
detected at other positions.

In Figure \ref{fig-02}, we overlaid the \hh\ \vone and \vtwo lines
from M82, as observed with the UTFPS. The \hh\ \vtwo\ emission at
(0, 0) is very similar to the line profile of the \vone\
transition while the match in line shapes is less clear at
(E14E\arcsec,N6\arcsec) where the signal to noise ratio of the
\vtwo\ observation is small. Fluxes and upper limits for the \hh\
\vtwo\ and \vthree\ lines toward NGC 253 and NGC 6946 are listed
in Table~\ref{tbl-3}.  Table~\ref{tbl-4} lists line ratios,
principally for the \vtwo\ to \vone\ ratio, along the major axis
of M82.  S/N for the ratio determination ranges from 2 to 6.
However, temporal variations in telluric absorption may increase
the uncertainty for the line ratios derived from the UTFPS data.

\subsection{ Velocity Integrated \hh\ Flux}

\subsubsection{ Fabry-Perot Spectrometer }

We determine the \hh\ line flux by integrating across the velocity
interval containing the line. We define the line emission
bandwidth, $\Delta V_L$, based on the $^{12}$CO \jone\ or \jtwo\
line spectra and list the values in Table~\ref{tbl-3}. For each
profile, we integrate over this velocity interval, subtracting a
continuum value determined from the average of the channels at the
two ends of the spectrum. We label the flux in the portion of the
spectrum where we are integrating over the line as ($f_{L,i}$) and
that over the spectral region containing only continuum as
($f_{C,i}$) where the subscript $i$ is an index for the individual
samples along the spectrum. The integrated line flux ($f$) is
calculated as:
\begin{equation}
  f = \Delta V_{ch} \sum_{i} \left( f_{L,i} - \bar{f_C} \right)\ ,
  \label{eq:flux}
\end{equation}
where $\bar{f_C}$ is the average of $f_{C,i}$ and $\Delta V_{ch}$
is the interval between the channels in the scanning mode (see
Table~\ref{tbl-2}). The error (or noise) of the integrated line
flux is derived from the standard deviation, $\sigma(f_{C,i})$, of
$f_{C,i}$:
\begin{equation}
  \sigma(f)
  = \Delta V_{ch} \sigma(f_{C,i})
    \sqrt{ N_L ( 1 + 1 / N_C ) }\ ,
    \label{eq:fluxsigma}
\end{equation}
where $N_L$ and $N_C$ are the numbers of individual spectral
samples in the emission and continuum bands, respectively.
Table~\ref{tbl-3} lists the observed positions and the measured
fluxes of the \hh\ \vone , \vtwo , and \vthree\ emission lines.
Note that the flux values, at the same wavelength and the same
position, vary with the beam size because the emission regions are
extended.

\subsubsection{ Long-slit Spectrometer }

The FWHM ($\Delta V_{FWHM} = 283$~\kms) of the instrumental
profile of CRSP is wider than the typical FWHM of M82 emission
lines. Assuming the line profile is gaussian, we derive the \hh\
line flux by using the single profile fitting task in the {\it
IRAF}\footnote{
    IRAF is distributed by the National Optical Astronomy
    Observatories, which are operated by the Association
    of Universities for Research in Astronomy, Inc.,
    under cooperative agreement with the National
    Science Foundation.}
package, SPLOT. The error in the flux is derived using
equation~(\ref{eq:fluxsigma}) with $1/N_C \approx 0$ and $\Delta
V_L$ $=$ $2 \times \Delta V_{FWHM}$.

\subsection{ Extinction Correction } \label{sec:Extinction}

The molecular hydrogen flux that we observe likely arises from the
surfaces of giant molecular clouds exposed to far-UV radiation
\citep{Pak96a,Pak96b,Davies03}. Therefore, as for the large-scale
CO line emission from galaxies, the area filling factor for the
observed \hh\ line emission is likely to be less than unity, so
that much of the emergent \hh\ will suffer only modest extinction
and the foreground screen model for extinction is appropriate.  If
the \hh\ emission is widespread and arises from the surfaces of
clouds, then there would only be a loss of half the \hh\ flux
(from the back side of each cloud), and a further loss from the
foreground extinction. The screen model is highly effective in
modeling the extended \hh\ line emission in the central kiloparsec
of the Milky Way. Here the \hh\ emission would not be detectable
at all if the emitting sources were uniformly mixed with molecular
gas clouds, since CO observations indicate these clouds typically
have \av\ $\simeq 1500$ \citep{Pak96a,Pak96b}.

The far-UV radiation that gives rise to the observed \hh\ line
emission arises from a slightly older population of stars than
those that excite \hii\ regions.  These older stars are not likely
to be still embedded in their natal clouds and can therefore
produce the observed widespread \hh\ emission. In contrast, the
\hi\ recombination lines likely emerge from \hii\ regions excited
by very early type stars that are still embedded in their natal
molecular clouds. Therefore, one might expect the extinction of
the \hi\ recombination lines to be significantly higher, and the
mixed extinction scenario may be more likely to apply to these
lines. However, the application of a screen model even to the
observed \brgamma\ flux from both ultraluminous IR galaxies
(ULIRGs) \citep{Goldader97b} and normal and nearby starburst
galaxies \citep{Kennicutt98} results in a reasonably tight
\brgamma\ to far-IR correlation that suggests no large extinction
corrections are required.

The foreground screen model is applicable for all of the galaxies
in our sample. The most controversial galaxy in our sample is the
nearby, highly inclined ($i = 82\degr$) galaxy, M82
\citep{Lynds63}. Extensive modeling of the observed H
recombination line emission from M82 suggests \av\ $=$ 3 to 9
magnitudes when a uniform foreground screen model is applied
\citep{Lester90,Puxley91,Satyapal97,Foerster01}. However, a model
where the dust and \hii\ regions are homogeneously distributed
also fits the data and suggests that \av\ $=$ 25 to 52 magnitudes
\citep{Puxley91}. We argue that while a mixed model {\it may}
apply for the \hi\ recombination lines, it is not appropriate for
the \hh\ lines due to the argument outlined above. To compare the
extinction corrected results, in Table~\ref{tbl-5}, we applied
both foreground model and mixed model to the \hh\ data of M82.

In converting from visual extinction, \av, in Table~\ref{tbl-5} to
IR extinction, $A_\lambda$, at different wavelengths, we used the
IR interstellar reddening law (Section 7.8. in \citet{Cox99} and
references therein):
\begin{equation}
    A_{\lambda} = 0.412\ A_V\ \lambda^{-1.75} \
    {\rm (valid \ for} \ 0.9 < \lambda < 6 \ \micron) ,
    \label{eq:Extinction}
\end{equation}
where $\lambda$ is in units of \micron. In
section~\ref{sec:H2Ratio} and Table~\ref{tbl-4}, where we compare
the \hh\ line ratios, we corrected for differential extinction for
lines at different wavelengths.

\subsection{ Intensity Distribution }   \label{sec:Distr}

Our new \hh\ measurements of galaxies are particularly useful for
comparison with other observations of line and continuum radiation
from neutral atomic and molecular gas because the size of our
UTFPS beam is more comparable to the sizes of beams used to
measure the other PDR tracers than are the arcsec-scale beam sizes
of typical near-IR slit spectrometers. Here we compare our \hh\
results to far-IR and submillimeter continuum, and \cii\
158~\micron\ and $^{12}$CO line data assembled from the
literature.

\subsubsection{ M82 } \label{sec:Distr_M82}

Figure~\ref{fig-09} shows the integrated intensities of \hh\
\vone, $^{12}$CO \jone\ \citep{Nakai87}, and $^{12}$CO \jfour\
lines \citep{Mao00} and the specific intensity of submillimeter
continuum at 450~\micron\ \citep{Hughes94} along the major axis of
M82 (P.A. $=$ $65\degr$; see Tables \ref{tbl-1} and \ref{tbl-3}).
The 450~\micron\ data were obtained with a 9\arcsec\ beam. We
smoothed these data to match the spatial resolution of our
16\arcsec\ beam observations. The beam sizes of the CO \jone\
\citep{Nakai87} and the CO \jfour\ \citep{Mao00} data are slightly
larger (16\arcsec\ and 18\arcsec, respectively) than the \hh\
beam.

In general, the distribution of CO, submillimeter continuum, and
\hh\ emission agree rather well.  The differences are consistent
with a common emission region with excitation differences
accounting for the small differences in source size. CO \jone\
observations show emission peaks on either side of the nucleus
(separation $\sim 26\arcsec$), indicating a rotating molecular
ring \citep{Shen95}. \citet{Mao00} observed a smaller separation
($\sim 15\arcsec$ or 270~pc) of the corresponding peaks in CO
\jfour. Thus, higher gas temperature and higher density clouds may
exist at the inner edges of the rotating molecular ring. While the
convolved scan map at 450~\micron\ that we show in
Figure~\ref{fig-09} has a plateau at the nucleus, the original
diffraction-limited map shows weak double peaks separated by $\sim
15\arcsec$ \citep{Hughes94}. This small separation supports the
idea that the rotating molecular ring has a temperature gradient,
since the dust continuum emission scales at least linearly with
temperature \citep{Sohn01}. The similarities in the CO,
submillimeter continuum, and \hh\ maps are consistent with a
picture in which the \hh\ emission from M82 arises primarily from
the molecular clouds in the circumnuclear ring.

\subsubsection{ NGC 253 }

Figure~\ref{fig-10} shows the integrated intensities of \hh\
\vone, $^{12}$CO \jone\ \citep{Canzian88}, and $^{12}$CO \jthree\
\citep{Israel95} as well as the specific intensity of
submillimeter continuum at 850~\micron\ \citep{Alton99} along the
molecular bar in NGC 253 which has an angular extent of $39\arcsec
\times 12\arcsec$ at a position angle of 64\degr. The molecular
bar is tilted from the major axis (P.A. = 51\degr,
\citealt{Pence80}) and apparently rotates as a rigid body
\citep{Canzian88}. The CO \jone\ data were obtained at the Owens
Valley Radio Observatory with a $5\arcsec \times 9\arcsec$ beam.
We smoothed them to match our UTFPS spatial resolution. The beam
sizes of the CO \jthree\ line and the 850~\micron\ continuum
observations are 14\arcsec\ and 15\arcsec, respectively.

The \hh\ emission distribution in Figure~\ref{fig-10} has a sharp
peak at the center and extends with relatively low intensities out
to $\sim 0.7$~kpc from the nucleus. The slight enhancement of \hh\
intensity at the southwestern bar (at the positions between $-300$
pc and $-100$ pc shown on the horizontal axis in
Figure~\ref{fig-10}) may be related to the southwest peak that is
also present in the overlaid CO \jone\ distribution. The
concentration of the \hh\ emission in this galaxy can be estimated
by comparing the flux measurements with different aperture sizes.
\citet{Engelbracht98} derived an \hh\ \vone\ flux from their
$2\farcs4 \times 12\arcsec$ slit of $(1.21 \pm 0.07) \times
10^{-16}$ \wattmeter. From our scanned observations along the bar
with the 11\arcsec\ beam ($100\arcsec \times 11\arcsec$), the
summed \hh\ \vone\ flux is $(2.91 \pm 0.34) \times 10^{-16}$
\wattmeter. From the 81\arcsec\ beam observations at the CTIO the
flux is $(3.17 \pm 0.25) \times 10^{-16}$ \wattmeter\ (see Tables
\ref{tbl-3} and \ref{tbl-5}). We conclude that the \hh\ emission
from the central $\sim 1$~kpc region is dominated by the
39\arcsec\ long molecular bar.

\subsubsection{ NGC 6946 }

$^{12}$CO $(J=1-0)$ observations of NGC 6946 show that the
molecular gas is concentrated in the central 10\arcsec\ region (or
0.3~kpc at the distance of 5.5~Mpc, \citealt{Ishizuki90a}). Using
near-IR long-slit spectroscopy, \citet{Engelbracht96} also found
that the \hh\ emission extends up to 12\arcsec\ along the major
axis. The flux in our 11\arcsec\ beam is $2.4 \times 10^{-17}$
\wattmeter. We expect that our beam covers most of the \hh\
emission in the central region of NGC 6946.

\subsubsection{ IC 342 }

An aperture synthesis map of $^{12}$CO \jone\ in the central
starburst region of IC 342 \citep{Ishizuki90b} shows a 7\arcsec\
(or 60 pc at the distance of 1.8~Mpc) diameter ring-like structure
within the extended ($\sim 40\arcsec$) molecular bar. Our beam
size, 16\arcsec\, covers the whole molecular ring structure. We
also observed regions offset by 16\arcsec\ along the molecular bar
(see Table~\ref{tbl-3}). Figure~\ref{fig-11} shows scan maps of
the integrated intensities of the \hh\ \vone, $^{12}$CO \jone\
\citep{Lo84}, and $^{12}$CO \jthree\ \citep{Steppe90} lines. Since
the spatial resolution of the CO \jone\ observations is 4\arcsec\
and that of the CO \jthree\ observations is 7\arcsec, we smoothed
the CO maps to match to our UTFPS spatial resolution of 16\arcsec.

In this galaxy, the \hh\ emission region is unresolved and
therefore significantly more compact than the CO emission region.
This difference between IC 342 and M82 and NGC 253 may, however,
be due to a more extreme difference in emissivities rather than to
a separation of emission regions. \citet{Boeker97} observed the
central region of IC 342 and found an \hh\ \vone\ emission region
corresponding to the 7\arcsec\ CO ring.

\subsection { Total Flux }

If \hh\ \vone\ line emission traces molecular gas illuminated by
far-UV photons, its distribution and strength should be related to
those of other PDR probes. We have compiled far-IR continuum, \oi\
line, and \cii\ line emission results from the literature. Far-IR
continuum, \oi, and \cii\ emission have been taken with low
spatial resolution ($0\farcm5 - 2\arcmin$). The individual beams
therefore cover $\sim 1$~kpc regions in the centers of nearby
galaxies ($d \approx 4$ Mpc). Our big beam Fabry-Perot data cover
comparable size scales.

\subsubsection { \hh\ \vone\ }

The CO interferometer map of M82 \citep{Shen95} shows that the
minor-axis width of the rotating ring-like bar is $\sim
10\arcsec$. Our scan map along the major axis with a 16\arcsec\
beam therefore most likely covers most of the \hh\ emission region
in the center. The total \hh\ \vone\ flux, $F_{H2}$, from the
$75\arcsec \times 16\arcsec$ region within our map is $(4.95 \pm
0.36) \times 10^{-16}$ \wattmeter\ (see Table~\ref{tbl-5}). NGC
253 is also an edge-on galaxy with the inclination of 78\degr\
\citep{Pence80}. The minor-axis width of the molecular bar, from
the CO observations, is $\sim 12\arcsec$ \citep{Canzian88}, which
is about the same as our 11\arcsec\ beam. For this galaxy, we have
the 81\arcsec\ CTIO measurement to provide the total \hh\ flux.
The active star forming regions in NGC 6946 and IC 342 are
relatively small compared to those in the starburst galaxies, M82
and NGC 253. One UTFPS beam covers the \hh\ emission region.

We converted the total \hh\ \vone\ flux values into extinction
corrected flux values assuming that the foreground model is
applicable (see Section~\ref{sec:Extinction} and
Table~\ref{tbl-5}).

\subsubsection{ Far-IR } \label{sec:L_FIR}

When making use of IRAS 60 \micron\ and 100 \micron\ observations,
one can define the total far-IR continuum flux, $F_{FIR}$, as the
flux measured from the source within a bandpass of 80 \micron\
centered at 82.5 \micron\ \citep{Helou88}. In addition to the IRAS
observations, the far-IR fluxes of the galaxies in our \hh\ sample
were measured using the University of Texas far-IR photometer (see
the references in \citealt{Smith96}) on the Kuiper Airborne
Observatory (KAO) and using the Long Wavelength Spectrometer (LWS;
\citealt{Clegg96}) on the Infrared Space Observatory (ISO;
\citealt{Kessler96}). The KAO observations spatially resolved the
emission regions with a beam size of $30\arcsec \times 40\arcsec$.
The spatial resolution of the IRAS data, which were processed
using a maximum correlation method algorithm \citep{Aumann90}, is
$100\arcsec \times 140\arcsec$ at 100 \micron, and that of the
ISO/LWS is $80\arcsec$ \citep{Gry01}.

In Section~\ref{sec:Distr}, we found that most of the
submillimeter dust continuum emission arises from regions $<
1\arcmin$ in diameter. Even though the measured luminosities from
the KAO instrument are slightly less than those from the IRAS and
ISO, the different aperture sizes do not significantly affect the
measured luminosity. Compiling the far-IR luminosities from the
IRAS, the KAO, and the ISO/LWS, we list in Table~\ref{tbl-6} the
average $F_{FIR}$ of the published values.

\subsubsection{ \oi\ 63 \micron\ and \cii\ 158 \micron }

The \oi\ 63 \micron\ flux, $F_{OI}$, and the \cii\ 158 \micron\
flux, $F_{CII}$ were measured with the U. C. Berkeley cryogenic
tandem Fabry-Perot spectrometer on the KAO (see the references in
\citealt{Stacey91}) and by the ISO/LWS. The FWHM beam sizes are
55\arcsec\ and 69\arcsec\ for the KAO and ISO respectively. Since
the emission regions are concentrated in the central $< 1\arcmin$
of the galaxies (Sections \ref{sec:Distr} and \ref{sec:L_FIR}),
the flux values measured by both KAO and ISO can be compared with
our \hh\ measurement of the galaxies. Table~\ref{tbl-6} lists the
\cii\ and \oi\ line fluxes.

\section{ DISCUSSION } \label{sec:Discussion}

A resolution of the long-running debate about the excitation
mechanism for the vibration-rotation \hh\ lines in galaxies and
about the physical nature of the line formation region is critical
if we are to reap any benefit from the diagnostic power of these
lines. We need to examine the observational evidence in the
context of models for the formation and excitation of \hh\ to
decide whether UV excitation followed by a radiative cascade or
collisional excitation account for the population of the upper
states of the near-IR transitions.  We need to determine if any
collisionally excited gas we do see was heated by UV light
incident on the neutral cloud or by mechanical energy in the form
of shocks.

Like giant molecular clouds in our own galaxy, the ISM of other
spirals and of starburst galaxies is bathed in UV radiation from
main-sequence OB stars that have already freed themselves from the
dense cores in which they formed.  Orion serves as a good example
for the galactic molecular clouds.  The \hh\ \vone\ emission
attributable to the PDRs produced by UV light has a luminosity of
34 \lsun\ over a 15 pc diameter region centered on $\theta$ Ori.
The luminosity of the same line attributable to the well-known and
very prominent Orion/KL shock is only 2.5 \lsun\ \citep{Luhman94}.
Observations of several \hh\ vibration-rotation transitions with
the UTFPS, including detection of the \vfive\ line, and
comparisons to other PDR tracers led \citet{Pak98} to conclude
that the majority of the \hh\ emission from several giant
molecular clouds in the Large Magellanic Cloud is a result of UV
excitation.

The ubiquity of UV emission incident on the surfaces of molecular
clouds in the inner regions of galaxies means that these galaxies
will always produce some UV-excited \hh\ emission.  The real
question then is whether this emission -- or emission from shocks
produced by some combination of protostellar jets, supernovae, and
large-scale gas flows -- will dominate the observed flux. In this
section, we examine whether UV excitation is sufficient to account
for all of the \hh\ emission observed toward the galaxies in our
sample.

\subsection{The \hh\ Line Ratios} \label{sec:H2Ratio}

The ratios of \hh\ lines arising in states with different upper
state energies and the \hh\ ortho- to para-ratios give us a way to
estimate the fraction of the emission produced by each excitation
process within the observed beam. In PDRs where $n(H_2)$ $<$ $5
\times 10^4$~cm$^{-3}$ (sources where a radiative cascade
determines the level populations), the ratios of \vtwo\ to \vone\
and \vthree\ to \vone\ are $\sim 0.6$ and $\sim 0.2$, respectively
\citep{Black87}. In shocked regions, e.g., bipolar outflows
\citep{Burton89,Davis99} and supernova remnants \citep{Richter95},
the observed ratio of \vtwo\ to \vone\ line flux is $\sim 0.1$,
which can be explained by a combination of J- and C-type bow
shocks \citep{Smith91,Smith94}. If the level populations of the
\hh\ molecules were in LTE, this ratio would be appropriate for
gas with $T_{exc} = 2000$~K \citep{Shull82,Black87}. At this
temperature, the ratio of \vthree\ to \vone\ is only $\sim 0.006$
(see Table~\ref{tbl-4}).

In Table~\ref{tbl-4}, the observed \hh\ ratios of \vtwo\ to \vone\
from M82, NGC 253, and NGC 6949 are $0.3-0.6$, which lies between
the ratios in the thermal and non-thermal excitation cases. If we
assume that \hh\ emission arises from a mixture of thermal
excitation at one LTE temperature and non-thermal excitation
following the rules for a UV-excited cascade, we can estimate the
fraction of the total emission in \hh\ lines contributed by
UV-excited gas.  In the discussion that follows, we use $\Re$ to
represent the fraction of the \hh\ emission (summed over all \hh\
transitions) contributed by UV fluorescence. Appendix~\ref{sec:Re}
gives a detailed explanation of how we derive $\Re$ from the
observed line ratios.

Figures \ref{fig-12}, \ref{fig-13}, and \ref{fig-14} show the
$\Re$ values derived from our UTFPS and CRSP data from M82
and NGC 253. We also plot the values of $\Re$ derived from other
observations of \hh\ emission in those galaxies
\citep{Foerster01,Harrison98,Engelbracht98}. The results show
that, at almost all positions in both galaxies, the majority of
the total \hh\ emission comes from regions excited radiatively by
UV photons. Figure \ref{fig-14} shows the values of $\Re$ derived
from a variety of \hh\ vibration-rotation transitions for NGC 253.
Most lines give values of $\Re$ between 0.6 and 1, again
indicating the predominance of UV-excited emission. If some of the
remaining (thermal) emission arises from dense gas in PDRs, the
non-thermal fraction may be even higher than the results from the
analysis used in the figures indicates. \citet{Sternberg89b}
modeled the \hh\ emission from warm PDRs, where the \hh\
vibrational level populations are thermalized by collisional
de-excitation. In these regions, the \hh\ line has excitation
temperature of $\sim 1000$~K. If we use the relative line
intensities derived assuming an excitation temperature, $T_{exc} =
1000$~K for the collisionally excited gas, the derived $\Re$
values are closer to unity than those derived assuming $T_{exc} =
2000$~K.

Our analysis shows that \hh\ emission from the central regions of
galaxies is mostly dominated by non-thermal excitation. UV
radiation may also be responsible for some of the remaining
thermal \hh\ emission if that emission arises in dense PDRs.  The
analysis presented in Appendix~\ref{sec:H2Fraction} and
illustrated in Figure~\ref{fig-15} points out that the picture one
gets from the \vone\ line alone is highly biased toward
collisionally excited emission, because this line's upper state
energy is lower than those of \vtwo\ or \vthree\ lines. A
UV-excited radiative cascade populates many rotational states in
higher vibrational levels. In purely radiatively excited \hh, the
\vone\ line accounts for only 1.6\% of the total \hh\ emission,
while the same line from 2000 K gas in LTE accounts for 8.5\% of
the ro-vibrational intensity \citep{Black87}. As a result, even
when only a small fraction of the \vone\ emission arises in
UV-excited regions, {\it most} of the overall \hh\ emission is a
product of UV excitation. In such cases, the thermal contribution
to the \vone\ emission could arise in dense PDR gas where UV
radiation is the ultimate source of the emission, or it could come
from shocks associated with supernova remnants or large-scale
flows.  A comparison of \hh\ \vone\ luminosity with fine structure
line emission from PDRs in galaxies will help distinguish between
these possibilities.

\subsection{\hh, \cii, and far-IR and the Origin of \hh\ Emission}
    \label{sec:H2CIIFIR}

In starburst nuclei, the far-IR luminosity arises mostly from PDRs
illuminated by recently formed OB stars.  Far-IR emission has been
used to estimate the star formation rates (SFRs) of such galaxies
\citep{Kennicutt98}. The $^2P_{3/2} \rightarrow ^2P_{1/2}$ C$^+$
emission at 158~\micron\ from galactic nuclei also arises mostly
from the warm photodissociated surfaces of far-UV exposed
molecular clouds, with lesser contributions from low density
ionized gas regions, and from the cold neutral medium
\citep{Stacey91}. The average \cii\ to far-IR luminosity ratio for
the four galaxies in our sample is $10^{-2.87 \pm 0.27}$ (see also
Figure~\ref{fig-16}), where the 0.27 represents the standard
deviation about the mean ratio. The fairly constant ratio is
consistent with a PDR origin for both \citep{Crawford85,Stacey91}.

In Figure~\ref{fig-16}, we also plot the ratio of \hh\ \vone\ to
far-IR luminosity versus the far-IR luminosity from our
observations. The $L_{H2}/L_{FIR}$ ratio is almost constant, with
a mean value of $10^{-5.05 \pm 0.28}$.
\citet{Goldader95,Goldader97a,Goldader97b} found a similar
proportionality between the extinction corrected \hh\ \vone\ and
far-IR luminosities from LIRGs ($10^{11}~L_\odot$ $<$ $L_{FIR}$
$<$ $10^{12.5}~L_\odot$). Their data give a value for
$L_{H2}/L_{FIR}$ from the LIRGs of $10^{-5.17 \pm 0.24}$, which is
consistent with our value for starbursts and normal spiral
galaxies. In calculating the mean ratio for nearby galaxies, we
assumed that the foreground model was applicable when making the
extinction correction to the M82 \hh\ data (see
Section~\ref{sec:Extinction}). If we applied the mixed model
result, 52 magnitudes of extinction, to the observed \hh\ \vone\
line flux, the corresponding ratio of \hh\ \vone\ line to far-IR
continuum of M82 would be very large, $L_{H2}/L_{FIR}$ $\simeq$
$10^{-2.9}$ (see Tables \ref{tbl-5} and \ref{tbl-6}). This ratio
is two orders of magnitude larger than the extinction corrected
line to continuum ratios typical for IR bright and LIRG galaxies
\citep{Goldader95, Burston01,Davies03} and six times larger than
ratio found for the most extreme \hh\ \vone\ luminous system, NGC
6240 \citep{Goldader95}.

The constant ratio of $L_{H2}/L_{FIR}$ and $L_{CII}/L_{FIR}$
suggests that the far-IR, \cii, and \hh\ emission regions and
their emission mechanisms are directly related. In examining this
possibility, we assume that the \hh\ \vone\ luminosity can
represent the total \hh\ transition luminosity. If the \hh\
emission is mostly from PDRs and its energy level populations are
partly thermalized, the relationship between the \hh\ \vone\ and
the total \hh\ transition luminosity is given by equations
(\ref{eq:Re_T}) and (\ref{eq:I_v}) in sections \ref{sec:Re} and
\ref{sec:H2Fraction}. When $\Re = 0.6$, 0.8, and 1, the ratio of
total \hh\ luminosity to luminosity in the I(1,0)S(1) line is 23,
34, and 63, respectively. Since the fraction of the non-thermal
excitation, $\Re$, observed in the galaxies in our sample is $0.8
\pm 0.2$, the error introduced by using the \vone\ line as an
indicator of the total \hh\ transition intensity is a factor of
$1.5-2$ (or $10^{0.2}-10^{0.3}$). This uncertainty is comparable
to the standard deviation of the observed $L_{H2}/L_{FIR}$ ratio
from normal and starburst galaxies ($10^{0.24}$), and LIRGs
($10^{0.28}$). Using the \vone\ line as a proxy for the total \hh\
luminosity, we now compare the observed \hh\ emission to the
intensity that independently derived PDR models would predict for
the same objects.

\subsection{The \hh\ to Far-IR Ratio Derived from PDR Models} \label{sec:model}

In this section, we examine the possibility that the observed \hh\
emission arises in PDRs by comparing the average observed \hh\
\vone\ intensity to the intensity predicted from PDR models with
parameters set by observed far-IR continuum and \oi\ and \cii\
line strengths. The emission from PDRs can be parameterized by the
cloud density, $n_H = n({\rm H}) + 2 n({\rm H_2})$, and the
strength of the far-UV radiation field illuminating the cloud,
$G_\circ$ \citep{Tielens85,Hollenbach99}. The fine structure line
ratio $F_{OI}/F_{CII}$ and the fine structure line to far-IR
continuum ratio $(F_{OI}+F_{CII})/F_{FIR}$ (see the values in
Table~\ref{tbl-6}) are relatively free from beam filling factor
effects and can be used to derive the density and local UV field
in the PDR. We compared the observed flux of \oi, \cii, and far-IR
(Table~\ref{tbl-6}) to the PDR models of \citet{Kaufman99}, and
list the best-fit values of $n_H$ and $G_\circ$ in
Table~\ref{tbl-7}. These values are different from the previous
results in \citet{Stacey91,Madden93,Carral94,Lord96}. One reason
for the difference is that the PDR model of \citet{Kaufman99} used
in this work includes additional gas heating due to polycyclic
aromatic hydrocarbons (PAHs) which results in a lower predicted
gas density \citep{Negishi01}.

The \hh\ \vone\ line intensity, $I_{H2}$, is also a function of
the strength of the external far-UV radiation field, $G_\circ$,
impinging on the PDR, and the gas density, $n_H$, in the PDR. In
order to compare this predicted \hh\ \vone\ intensity for a
derived $n_H$, $G_\circ$ pair with the observed values, e.g.,
$L_{H2}/L_{FIR}$ in Figure~\ref{fig-16}, it is necessary to derive
the intensity ratio of \hh\ \vone\ to far-IR, $I_{H2}/I_{FIR}$
from the models. We can convert the derived far-UV radiation field
strength, $G_\circ$ in Table~\ref{tbl-7}, into far-IR intensity,
$I_{FIR}$, assuming that most of the far-UV photons are absorbed
by grains and reradiated in the far-IR:
\begin{equation}
  I_{FIR} = 2 \times ( 1.3 \times 10^{-7} )\ G_\circ \
    {\rm W~m^{-2}~sr^{-1}}\ , \label{eq:G0_to_IFIR}
\end{equation}
where the factor of 2 is for the photons with $\lambda >
206.6~{\rm nm}$. Figure~\ref{fig-17} shows the intensity ratio,
$I_{H2}/I_{FIR}$, as a function of $G_\circ$ and $n_H$. To
calculate the \hh\ intensity, we used a second PDR model that
calculates abundances and temperatures in a plane-parallel slab
for a given $n_H$ and $G_\circ$ \citep{Black87,Jansen95} and uses
a Monte Carlo method for the radiative transfer \citep{Choi95}. We
assume solar metallicity in the models because the \hh\ intensity
is not sensitive to the metallicity \citep{Pak98}.

In Figure~\ref{fig-17}, we plot the locations of the galaxies in
our sample in ($G_\circ$, $n_H$) space, as derived from the \oi\
and \cii\ observations using the PDR models. The figure also shows
the model predictions for the I$_{H2}/I_{FIR}$ ratio.
Table~\ref{tbl-7} lists the values, as well as the predicted
strengths of the \hh\ \vone\ line. Since the derived $n_H$ values
of NGC 6946 and NGC 253 are lower than the model domain, we
extrapolated the grids. The resulting values of $I_{H2}/I_{FIR}$
are listed in Table~\ref{tbl-7}. The average of the \hh\ to far-IR
ratios predicted by the PDR models from the fine-structure line
observations is $10^{-5.13 \pm 0.36}$ which is in very close
agreement with the value of $10^{-5.05 \pm 0.28}$ measured
directly (see Section~\ref{sec:H2CIIFIR}). This agreement argues
strongly for UV excitation in PDRs as the primary source of \hh\
\vone\ and, by implication, as the dominant cause of \hh\ emission
overall in the central regions of galaxies.  Our results are in
good agreement with those of \citet{Davies03} who conclude that
the observed \hh\ line emission from a sample of seven ULIRGs
arises from dense PDRs exposed to strong ($G_\circ$ $\simeq$
$10^3$) radiation fields.

\subsection{ Reasons for the Constant \hh\ to Far-IR Ratio}

In our study of the inner 400~pc of the Milky Way, we found that
most \hh\ emission arises as a result of UV excitation and that
the ratio of \hh\ \vone\ luminosity to far-IR luminosity is $\sim
10^{-4.8}$ \citep{Pak96a,Pak96b}. As shown in
sections~\ref{sec:H2Ratio}, \ref{sec:H2CIIFIR}, and
\ref{sec:model}, the results for other galaxies are similar; UV
excitation dominates and the $L_{H2}/L_{FIR}$ ratios are
$10^{-5.05 \pm 0.28}$ for spiral galaxies and local starburst
galaxies and $10^{-5.17 \pm 0.24}$ for LIRGs.

The relationship between far-IR intensity and \hh\
intensity as conditions in a PDR vary can be explained in terms of
the competition between \hh\ absorption and dust absorption of
far-UV photons. \citet{Burton90} showed analytically that the
penetration of far-UV photons into a cloud is determined by \hh\
self-shielding rather than by dust extinction when:
\begin{equation}
  \frac{n_H}{G_\circ} > 80\ {\rm cm^{-3}}. \label{eq:absorption}
\end{equation}
In Figure~\ref{fig-17}, the dotted line divides the ($G_\circ$,
$n_H$) plane into the \hh\ self-shielding dominant region and the
region where shielding by dust is the major protector of \hh. In
the \hh\ self-shielding case, the UV-excited \hh\ intensity,
$I_{H2}$, is proportional to the far-UV field strength, $G_\circ$
because any additional UV photons are absorbed by \hh\ and most of
these absorptions result in a fluorescent cascade.  In this case,
the $I_{H2}/I_{FIR}$ ratio is always $\sim 10^{-4}$. In the case
where dust shielding dominates, the \hh\ intensity is independent
of the far-UV field but scales with the gas density, $n_H$ and
the $I_{H2}/I_{FIR}$ ratio is $< 10^{-4}$. The distribution of our
sample galaxies in the ($G_\circ$, $n_H$) plane
(Figure~\ref{fig-17}) shows that the ratio of $n_H/G_\circ$,
derived from the observations of \cii\ and \oi\ lines and far-IR
continuum (see section~\ref{sec:model}), is lower than 80 in all
cases, consistent with the observed $I_{H2}/I_{FIR}$ ratio. Dust
absorption of far-UV photons should therefore dominate over \hh\
self-shielding in the central regions of these galaxies.

The observed constant ratio of \hh\ \vone\ intensity to far-IR
intensity is surprising in view of the results of the PDR
modelling. In the dust-shielding regime, this constant ratio can
only come about if the gas density increases as the far-UV field
strength increases. For a uniform medium, with an observed
$\log(I_{H2}/I_{FIR})$ $=$ $-5$, the implied relationship is
\begin{equation}
  n_H \propto G_\circ ^{1.3} \ . \label{eq:MyWork}
\end{equation}

This inferred relation joins a long string of similar correlations
between mass or density and surface density indicators and proxies
for the star formation rate.  These include: the Schmidt law
\citep{Schmidt59}, as determined from H$\alpha$ and CO
J=$1\rightarrow 0$ observations \citep{Kennicutt98}, the scaling
of $n_H$ and G$_\circ$ derived from \oi, \cii\ and far-IR
intensities \citep{Malhotra01}, the strong correlation of the
bolometric luminosity and virial mass of dense galactic cores
derived from CS and far-IR observations \citep{Shirley03} and the
close correlation of bolometric luminosity with the amount of
dense gas in galaxies \citep{Solomon92}. One possibility, then, is
that the relationship between density and incident UV field
inferred from the constant $I_{H2}/I_{FIR}$ ratio is telling us
that the surfaces of molecular clouds in galaxies, after stars
have begun to form, have higher densities when the local UV field
is higher. This conclusion is not unreasonable, given that less
dense clumps will be ionized and photo-evaporated more rapidly and
that the average densities of surviving clumps should be higher
when the incident UV field is stronger.

\section{CONCLUSIONS} \label{sec:Conclusions}


We have analyzed near-IR \hh\ emission from the central $\sim
1$~kpc regions in the nearby starburst galaxies, M82 and NGC 253,
and the vigorously star-forming galaxies, NGC 6946 and IC 342. The
line ratios of various \hh\ vibration-rotation lines, e.g., \vtwo,
and \vonez\ to \vone, show that, while both thermal and
non-thermal excitation are responsible for the emission,
non-thermal excitation predominates. In addition to the \hh\ line
ratios, the constant ratio between \hh\ \vone, \cii, and far-IR
luminosity suggests that the emission in the two lines and the
far-IR continuum originate together in photodissociation regions.

We also compared scan maps of \hh\ \vone\ emission along the major
axes (or molecular bars) of the galaxies with maps of CO and
sub-mm emission. The distribution of \hh\ line emission correlates
with the sub-mm continuum emission, and also agrees well with the
distribution of CO emission. The \hh\ emission traces active star
formation regions where the far-UV fields are intense and the gas
is dense and warm. The correlation of \hh\ and sub-mm emission
suggests that clouds have similar structure or that the star
formation and cloud distributions are similar in the central
regions of these galaxies.

We compare the luminosities of \hh\ \vone, \cii\ 158 \micron, and
far-IR continuum. The $L_{H2}/L_{FIR}$ ratios of normal spiral and
starburst galaxies are the same. Comparing the observed ratio to
PDR models, we conclude that the column of warm \hh\ line emitting
gas is limited by the extinction of far-UV photons by dust, rather
than by \hh\ self-shielding in the PDRs. In this case, the \hh\
intensity is only sensitive to the gas density. The constant
$L_{H2}/L_{FIR}$ ratio over the large far-IR luminosity ranges of
the galaxies, then requires that the gas density in the PDRs
increase as the far-UV (or far-IR) intensity increases.



We thank Mark R. Swain for help with observations at the Palomar
Observatory, Ewine van Dishoeck for providing PDR model results,
and Michael Luhman and Thomas Benedict for their work on the
UTFPS. This work was supported by NSF Grant 95-30695. S.P.
appreciates the warm hospitality of the staff of the Institute of
Space and Astronautical Science in Japan.

\appendix

\section{Derivation of Total \hh\ Emission} \label{sec:derivation}

\subsection{Fraction of Non-Thermal Excitation of \hh }
\label{sec:Re}

Assuming that \hh\ emission arises from a mixture of thermal and
non-thermal excitation, the total \hh\ intensity summing over all
vibration-rotation transitions, $I$, and the intensity of a single
transition, $I_v$, can be expressed as
\begin{eqnarray}
  I &=& I^{th} + I^{nth} = \sum_v I_v \ ,\ {\rm and} \\
  I_v &=& I_v^{th} + I_v^{nth} \ ,
\end{eqnarray}
where the superscripts $th$ and $nth$ refer to the thermal
and non-thermal excitation, respectively. We can define the
fraction of non-thermal \hh\ intensity in the observed \hh\
intensity as
\begin{eqnarray}
  \Re &=& \frac{ I^{nth} }{ I } \ ,\ {\rm and}  \\
  \Re_v &=& \frac{ I_v^{nth} }{ I_v } \ .
\end{eqnarray}
The fractional intensity of thermal or non-thermal emission in
a single transition relative to the total thermal or non-thermal
emission is:
\begin{eqnarray}
  s_v &=& \frac{ I_v^{th} }{ I^{th} } \ ,\ {\rm and}  \label{eq:s_v} \\
  p_v &=& \frac{ I_v^{nth}}{ I^{nth}} \ . \label{eq:p_v}
\end{eqnarray}
Table~\ref{tbl-8} lists the adopted values of $s_v$ and $p_v$ for
\hh\ \vone, \vtwo, \vthree, and \vonez\ from model calculations
\citep{Black87}. We assumed that the excitation temperature is
2000~K for thermalized gas. From the above equations we can derive
the \hh\ line ratio as
\begin{equation}
  \frac{I_v}{I_1}
  = \frac{ s_v - (s_v - p_v) \Re }
  { s_1 - (s_1 - p_1) \Re } \ ,     \label{eq:H2ratio}
\end{equation}
where  the subscript `1' denotes the \vone\ transition line.
From equation (\ref{eq:H2ratio}), we can rewrite, $\Re$, as a
function of the observed \hh\ ratio of another transition to the
\vone\ line:
\begin{equation}
  \Re = \frac{ s_v - s_1 (I_v/I_1) }
        { (s_v-p_v) - (s_1-p_1) (I_v/I_1) } \ .     \label{eq:Re_T}
\end{equation}

\subsection{ Fraction of \hh\ \vone\ Line Intensity }
\label{sec:H2Fraction}

The \hh\ \vone\ at 2.1218
\micron\ line is the most commonly observed near-IR \hh\ line
because the wavelength resides in a clear part of the atmospheric
window and the line is usually brighter than other \hh\ lines. When the
level populations of excited \hh\ molecules are partly
thermalized, the \vone\ emission is enhanced relative to most
other \hh\ lines because of its lower excitation requirements. We
need to check whether the \hh\ \vone\ emission can represent the
total \hh\ intensity.

We can generalize the fraction of the non-thermal excitation in
a single transition, $\Re_v$, in terms of $\Re$ as:
\begin{equation}
  \Re_v = \left[
        1 - \frac{s_v}{p_v} \left( 1 - \frac{1}{\Re} \right)
        \right]^{-1} \ .
        \label{eq:Re_v}
\end{equation}
Figure~\ref{fig-15} shows $\Re_v$ versus $\Re$ for various
transitions. For example, when $\Re \simeq 0.6$, about 20 percent
of the \hh\ \vone\ emission arises from non-thermal excitation. On
the other hand, we can extrapolate to the total \hh\
vibration-rotation transition intensity, $I$, from a single \hh\
transition line emission, $I_v$, as
\begin{equation}
  I = \frac{I_v}{ s_v - ( s_v - p_v ) \Re } \ . \label{eq:I_v}
\end{equation}


\clearpage
\begin{figure}
   \caption{
   M82 spectra of the \hh\ \vone\ line observed
with the UTFPS of a 16\arcsec\ beam (solid lines). The dotted
lines show $^{12}$CO \jone\ spectra \citep{Nakai87} taken with the
16\arcsec\ beam of the Nobeyama 45~m telescope and smoothed to the
spectral resolution of the \hh\ spectra. The R.A. and Dec. offsets
in the upper left corner of each spectrum are in arc seconds from
the center of the galaxy (see Tables~\ref{tbl-1} and \ref{tbl-3}).
The Y axis gives the intensity scale for the \hh\ spectra while
the intensity of each CO spectrum has been scaled to match the
corresponding \hh\ spectrum. The two thick vertical ticks on the X
axis mark the emission bandwidth, $\Delta V_L$, within which we
integrated the emission line fluxes (see Table~\ref{tbl-3}).
   \label{fig-01}
   }
\end{figure}

\begin{figure}

   \caption{
   M82 spectra of the \hh\ \vone\ (left panel) and \vtwo\
(right panel) lines observed with the UTFPS of a 11 \arcsec\ beam.
The dotted lines show CO \jone\ spectra taken with a 16 \arcsec\
beam (see Fig.~\ref{fig-01}).
   \label{fig-02}
   }
\end{figure}

\begin{figure}
   \caption{
   NGC 253 spectra of the \hh\ \vone\ line observed
with the UTFPS of a 11\arcsec\ beam (solid lines). We have
overlaid $^{12}$CO \jtwo\ observations obtained by
\citet{Mauersberger96} (beam size 12\arcsec, dotted lines),
smoothed in the spectral direction to the resolution of the \hh\
spectrum). See Fig.~\ref{fig-01} for further explanation.
   \label{fig-03}
   }
\end{figure}

\begin{figure}
   \caption{
   NGC 253 spectra of the \hh\ \vone\ (left panel) and \vtwo\
(right panel) lines observed with the UTFPS of a 11 \arcsec\ beam.
See Figures~\ref{fig-01}, \ref{fig-02}, and \ref{fig-03} for
further explanation.
   \label{fig-04}
   }
\end{figure}

\begin{figure}
   \caption{
   NGC 253 spectra of \hh\ \vone\ (left panel), \vtwo\ (middle panel),
and \vthree\ (right panel) lines observed toward the center of the
galaxy with the UTFPS of an 81\arcsec\ beam. We have overlaid
$^{12}$CO \jone\ observations obtained by \citet{Paglione01} with
the 45\arcsec\ beam of the FCRAO 14~m telescope, smoothed to the
spectral resolution of the \hh\ observations.
   \label{fig-05}
   }
\end{figure}

\begin{figure}
   \caption{
   NGC 6946 spectra of \hh\ \vone\ (left panel) and \vtwo\ (right panel)
lines observed toward the center of the galaxy with the UTFPS of
an 11\arcsec\ beam. The dotted line shows $^{12}$CO \jone\
observations with the 16\arcsec\ beam \citep{Sofue88}.
   \label{fig-06}
   }
\end{figure}

\begin{figure}
   \caption{
   IC 342 spectra of \hh\ \vone\ line (solid line) obaserved
with the UTFPS of a 16\arcsec\ beam. We smoothed the $^{12}$CO
\jtwo\ observations (dotted line; \citealt{Xie94}), taken with a
23\arcsec\ beam, to the spectral resolution of the \hh\ results.
   \label{fig-07}
   }
\end{figure}

\begin{figure}
   \caption{
   M82 spectra observed with the long slit spectrometer, CRSP.
Each spectrum shows the average over a $5\farcs6 \times 5\farcs5$
box. The positions of the \hh\ \vone, \vthree, \vonez, and \vtwo\
lines are marked by arrows. Offsets for the spectra are in arc
seconds from the central position given in Table~\ref{tbl-1}.
   \label{fig-08}
   }
\end{figure}

\begin{figure}
  \caption{
  Intensity distributions of
  \hh\ \vone\ (solid line; this work),
  450 \micron\ continuum (dotted line; Hughes et al. 1994),
  $^{12}$CO \jone\ (dashed line; Nakai et al. 1987), and
  $^{12}$CO \jfour\ (dash-dotted line; Mao et al. 2000)
  from M82 along the major axis (P.A. $=$ $65\degr$).
Positive numbers along the horizontal axis increase to the
northeast on the sky. We convolved the 450 \micron\ data to match
the size of the \hh\  and CO beams. The ordinate is in units of integrated
\hh\ intensity. The intensity levels of the other tracers
are normalized to match the \hh\ peak. The error bars for the \hh\
data give the intensity uncertainty at each point.
  \label{fig-09}
  }
\end{figure}

\begin{figure}
  \caption{
  Intensity distribution along the molecular bar of NGC 253 of
  \hh\ \vone\ (solid line; this work), compared to the
850 \micron\ continuum (smoothed to 16 \arcsec\ resolution, dotted
line; Alton et al. 1999), $^{12}$CO \jone\ (dashed line; Canzian
et al. 1988), and $^{12}$CO \jthree\ (dash-dotted line; Israel et
al. 1995).
  \label{fig-10}
  }
\end{figure}

\begin{figure}
  \caption{
  Intensity distributions of
  \hh\ \vone\ (solid line; this work),
  $^{12}$CO \jone\ (dashed line; Lo et al. 1984), and
  $^{12}$CO \jthree\ (dash-dotted line; \citealt{Steppe90})
  along the IC 342 molecular bar (P.A. $=$ $14\degr$).
  \label{fig-11}
  }
\end{figure}

\begin{figure}
    \caption{
    Distribution of the derived ratio of non-thermal to total
ro-vibrational \hh\ emission (see eq.~[\ref{eq:Re_T}]), $\Re$,
based on the observed ratios of \hh\ \vtwo\ to \vone\ (top plot)
and \vonez\ to \vone\ (bottom plot) in M82. The large filled
circles denote $\Re$ values derived from UTFPS 11\arcsec\ beam
data. The small filled circles and the small filled squares denote
$\Re$ derived from CRSP $5\farcs5 \times 5\farcs6$ aperture data.
The open diamonds and the open squares denote $\Re$ derived from
line ratios obtained with a $2\farcs5 \times 2\farcs5$ (or 44~pc
$\times$ 44~pc) aperture by \citet{Foerster01}. The abscissa shows
the projected positions from the center along the major axis (P.A.
of 65\degr\ for the UTFPS data and the \citet{Foerster01} data and
73\degr\ for the CRSP data).
    \label{fig-12}
}
\end{figure}

\begin{figure}
    \caption{
    Distribution of $\Re$ derived from the observed ratios
of \hh\ \vtwo\ to \vone\ (top plot) and \vonez\ to \vone\ (bottom
plot) in NGC 253. The abscissa shows the projected positions from
the center along the major axis (P.A. = 51\degr). The filled
circle denotes $\Re$ derived using the UTFPS 11\arcsec\ beam data.
The open diamonds and the open squares denote show values derived
from spectra taken with a $3\arcsec \times 3\arcsec$ (or 36~pc
$\times$ 36~pc) aperture by \citet{Harrison98}.
    \label{fig-13}
    }
\end{figure}

\begin{figure} \epsscale{0.65}
    \caption{
    $\Re$ values derived from the observed ratios of
various \hh\ lines to \vone\ line in NGC 253. The filled circles
denote the UTFPS 81\arcsec\ (or 980~pc) beam data from this work.
The open squares denote the grating spectrometer data with a
$2\farcs4 \times 12\arcsec$ (or 29~pc $\times$ 140~pc) aperture by
\citet{Engelbracht98}. All data in this plot were observed toward
the center of the galaxy. The abscissa shows the wavelength of the
\hh\ line that was compared to the \vone\ line.
    \label{fig-14}
    }
\end{figure}

\begin{figure}
    \caption{
    Plot of $\Re_v$ versus $\Re$ using equation~(\ref{eq:Re_v}).
$\Re_v$ is the fraction of non-thermal \hh\ emission intensity in
the particular line (e.g., \vone, \vonez, \vtwo, or \vthree)
indicated against the plotted curve.
    \label{fig-15}
    }
\end{figure}

\begin{figure}
  \caption{
   Ratios of \cii\ and \hh\ \vone\ line luminosity
to far-IR continuum luminosity toward the galaxies in our sample.
The \cii\ and \hh\ luminosity ratios are denoted by the open
diamonds and the filled circles, respectively. See
Table~\ref{tbl-6} for references and a detailed
explanation.
   \label{fig-16}
   }
\end{figure}

\begin{figure}
   \caption{
   Ratios of \hh\ \vone\ line intensity to far-IR continuum intensity as
a function of the gas density, $n_H$, and the far-UV strength,
$G_\circ$. The contour labels are in log scale. The dotted line
divides the ($G_\circ$, $n_H$) plane into the \hh\ self-shielding
dominant region and the dust absorption dominant region (see
equation~[\ref{eq:absorption}]). The sample galaxies, whose
$G_\circ$ and $n_H$ are derived from the \cii\ and \oi\
observations, are overlayed.
   \label{fig-17}
   }
\end{figure}


\clearpage
\begin{deluxetable}{lccccccc}
    \tablewidth{0pt}
    \tabletypesize{\footnotesize}
    \tablecaption{ Adapted Parameters for Galaxies   \label{tbl-1} }
    \tablehead{
\colhead{   Galaxy  } &  \colhead{  $\alpha (J2000)$    } &  \colhead{  $\delta (J2000)$    } &  \colhead{  ref } &  \colhead{  P.A.    } &  \colhead{  ref } &  \colhead{  Distance    } &  \colhead{  ref } \\
\colhead{       } &  \colhead{  ($^h$ $^m$ $^s$)    } & \colhead{
(\arcdeg\ \arcmin\ \arcsec )    } &  \colhead{      } &  \colhead{
(\arcdeg )  } &  \colhead{      } &  \colhead{ (Mpc)   } &
\colhead{      }
    }
\startdata
    M82 &   09 55 52.6  &   $+69$ 40 47 &   1   &   65\tablenotemark{a} &   2   &   3.6 &   3   \\
    NGC 253 &   00 47 33.1  &   $-25$ 17 16 &   4   &   64\tablenotemark{b} &   5   &   2.5 &   6   \\
    NGC 6946    &   20 34 52.2  &   $+60$ 09 14 &   7   &   62\tablenotemark{a} &   8   &   5.5 &   9   \\
    IC 342  &   03 46 48.1  &   $+68$ 05 46 &   10  &   14\tablenotemark{b} &   11  &   1.8 &   12  \\
\enddata
    \tablenotetext{a}{Position angle of the major axis.}
    \tablenotetext{b}{Position angle of the molecular bar.}
\tablerefs{
    (1) \citealt{Rieke80};
    (2) \citealt{Nilson73};
    (3) \citealt{Freedman94};
    (4) \citealt{Becklin73};
    (5) \citealt{Canzian88};
    (6) \citealt{Mauersberger96} (references therein);
    (7) \citealt{Turner83};
    (8) \citealt{Rogstad72};
    (9) \citealt{Ishizuki90a} (references therein);
    (10) \citealt{Becklin80};
    (11) \citealt{Ishizuki90b};
    (12) \citealt{Boeker99} (references therein)
}
\end{deluxetable}

\begin{deluxetable}{lcccccccl}
    \tablewidth{0pt}
    \tabletypesize{\scriptsize}
    \tablecaption{ Observation Logs and Instrument Parameters for
    Fabry-Perot Spectrometer
    \label{tbl-2} }
    \tablehead{
\colhead{   Galaxy  } &  \colhead{  \hh\ Line\tablenotemark{a}  } &  \colhead{  Observation } &  \colhead{  Observatory\tablenotemark{b}    } &  \colhead{  $\theta_{ED}$\tablenotemark{c}  } &  \colhead{  $\Delta V_{FWHM}$\tablenotemark{d}  } &  \colhead{  $\Delta V_{BW}$\tablenotemark{e}    } &  \colhead{  $\Delta V_{ch}$\tablenotemark{f}    } &  \colhead{  Flux Calibration    } \\
\colhead{       } &  \colhead{  (\micron)   } &  \colhead{ Date }
&  \colhead{      } &  \colhead{  (\arcsec )  } &
\multicolumn{3}{c}{            (\kms )         } &  \colhead{ Star
}
    }
\startdata
    M82 &   \vone\ 2.12183  &   1995 Dec    &   Palomar &   16    &   100 &   156 &   43.8    &   HR 8505 \\
        &       &   1996 Sep    &   Palomar &   11    &   105 &   164 &   35.0    &   HR 7924 \\
        &   \vtwo\ 2.24771  &   1996 Sep    &   Palomar &   11    &   111 &   173 &   35.0    &   HR 8781 \\
    NGC 253 &   \vone\ 2.12183  &   1995 Oct    &   CTIO    &   81  &   125 &   192 &   43.8    &   HR 6084, 7525, 8728 \\
        &       &   1996 Sep    &   Palomar &   11    &   105 &   164 &   35.0    &   HR 7924 \\
        &   \vtwo\ 2.24771  &   1995 Oct    &   CTIO    &   81  &   132 &   203 &   35.0    &   HR 6084 \\
        &       &   1996 Sep    &   Palomar &   11    &   111 &   173 &   35.0    &   HR 8781 \\
        &   \vthree\ 2.20139    &   1995 Oct    &   CTIO    &   81  &   129 &   198 &   43.8    &   HR 6084, 8728   \\
    NGC 6946    &   \vone\ 2.12183  &   1996 Sep    &   Palomar &   11    &   105 &   164 &   35.0    &   HR 7924 \\
        &   \vtwo\ 2.24771  &   1996 Sep    &   Palomar &   11    &   111 &   173 &   35.0    &   HR 8781 \\
    IC 342  &   \vone\ 2.12183  &   1995 Dec    &   Palomar &   16    &   100 &   156 &   25.0    &   HR 8505 \\
\enddata
    \tablenotetext{a}{Rest wavelengths in vacua.}
    \tablenotetext{b}{We used the CTIO 1.5 m telescope and the Palomar 5 m telescope.}
    \tablenotetext{c}{Diameter of the equivalent disk. See the text for details.}
    \tablenotetext{d}{Full-width at half-maximum of the instrument spectral profile for an extended source.}
    \tablenotetext{e}{Band width of the instrument profile. $\Delta V_{BW} = \int I(V) dV / I_{peak}$ }
    \tablenotetext{f}{Step distance between the wavelength channels in the scanning mode.}
\end{deluxetable}

\begin{deluxetable}{lrrrrrrrr}
    \tablewidth{0pt}
    \tabletypesize{\scriptsize}
    \tablecaption{ Observed Flux of H$_2$ Lines   \label{tbl-3} }
    \tablehead{
\colhead{   Galaxy  } &  \multicolumn{2}{c}{    $\theta_{ED}$           } &  \multicolumn{2}{c}{    Position            } &  \colhead{  $\Delta V_{L}$\tablenotemark{b} } &  \colhead{  \vone           } &  \colhead{  \vtwo           } &  \colhead{  \vthree         } \\
\colhead{       } &  \colhead{      } &  \colhead{      } &  \colhead{  $\Delta \alpha$\tablenotemark{a}    } &  \colhead{  $\Delta \delta$\tablenotemark{a}    } &  \colhead{      } &  \colhead{  2.12183 \micron         } &  \colhead{  2.24771 \micron         } &  \colhead{  2.20139 \micron         } \\
\colhead{       } &  \colhead{  (\arcsec )  } &  \colhead{  (pc)
} &  \multicolumn{2}{c}{    (\arcsec )          } &  \colhead{
(\kms ) } &  \multicolumn{3}{c}{    ($10^{-17}$ W m$^{-2}$)
}
    }
\startdata
    M82 &   11.1    &   194 &   14  &   6   & $ +80, +430   $ & $   4.00    \pm 1.04    $ & $   2.61    \pm 0.73    $ & \nodata         \\
        &       &       &   0   &   0   & $ +80, +430   $ & $   11.79   \pm 2.05    $ & $   9.28    \pm 0.87    $ & \nodata         \\
        &   15.7    &   274 &   41  &   19  & $ +130, +480  $ & $   -1.10   \pm 0.81    $ & \nodata         &   \nodata         \\
        &       &       &   27  &   13  & $ +130, +480  $ & $   3.74    \pm 0.39    $ & \nodata         &   \nodata         \\
        &       &       &   14  &   6   & $ +80, +430   $ & $   13.12   \pm 1.71    $ & \nodata         &   \nodata         \\
        &       &       &   0   &   0   & $ +80, +430   $ & $   15.99   \pm 2.51    $ & \nodata         &   \nodata         \\
        &       &       &   -14 &   -6  & $ 0, +350 $ & $   12.09   \pm 1.64    $ & \nodata         &   \nodata         \\
        &       &       &   -27 &   -13 & $ 0, +350 $ & $   4.51    \pm 1.03    $ & \nodata         &   \nodata         \\
        &       &       &   -41 &   -19 & $ 0, +350 $ & $   -1.10   \pm 1.19    $ & \nodata         &   \nodata         \\
\\
    NGC 253 &   11.1    &   135 &   45  &   22  & $ +40, +360   $ & $   0.53    \pm 1.52    $ & \nodata         &   \nodata         \\
        &       &       &   36  &   18  & $ +40, +360   $ & $   2.81    \pm 0.79    $ & \nodata         &   \nodata         \\
        &       &       &   27  &   13  & $ +40, +360   $ & $   0.71    \pm 1.09    $ & \nodata         &   \nodata         \\
        &       &       &   18  &   9   & $ +40, +360   $ & $   1.62    \pm 1.09    $ & \nodata         &   \nodata         \\
        &       &       &   9   &   4   & $ +40, +360   $ & $   2.05    \pm 0.86    $ & \nodata         &   \nodata         \\
        &       &       &   0   &   0   & $ +70, +390   $ & $   9.71    \pm 0.91    $ & $   3.53    \pm 1.54    $ & \nodata         \\
        &       &       &   -9  &   -4  & $ +100, +420  $ & $   6.87    \pm 0.79    $ & \nodata         &   \nodata         \\
        &       &       &   -18 &   -9  & $ +100, +420  $ & $   2.69    \pm 1.04    $ & \nodata         &   \nodata         \\
        &       &       &   -27 &   -13 & $ +100, +420  $ & $   1.22    \pm 0.89    $ & \nodata         &   \nodata         \\
        &       &       &   -36 &   -18 & $ +100, +420  $ & $   0.03    \pm 1.22    $ & \nodata         &   \nodata         \\
        &       &       &   -45 &   -22 & $ +100, +420  $ & $   0.86    \pm 0.73    $ & \nodata         &   \nodata         \\
        &   81  &   982 &   0   &   0   & $ +70, +390   $ & $   31.7    \pm 2.5 $ & $   20.9    \pm 9.2 $ & $   5.8 \pm 2.9 $ \\
\\
    NGC 6946    &   11.1    &   296 &   0   &   0   & $ -60, +250   $ & $   2.42    \pm 0.53    $ & $   0.75    \pm 0.40    $ & \nodata         \\
\\
    IC 342  &   15.7    &   137 &   5   &   15  & $ -70, +120   $ & $   -0.11   \pm 0.37    $ & \nodata         &   \nodata         \\
        &       &       &   0   &   0   & $ -70, +120   $ & $   1.71    \pm 0.39    $ & \nodata         &   \nodata         \\
        &       &       &   -5  &   -15 & $ -70, +120   $ & $   0.13    \pm 0.46    $ & \nodata         &   \nodata         \\
\enddata
    \tablenotetext{a}{Offset from the centers listed in Table~\ref{tbl-2}. }
    \tablenotetext{b}{Emission line range (minimum, maximum) for the integrated line flux. See equation~(\ref{eq:flux}).}
\end{deluxetable}

\begin{deluxetable}{lcccccc}
    \tablewidth{0pt}
    \tabletypesize{\scriptsize}
    \tablecaption{ Observed H$_2$ Line Ratios\tablenotemark{a}  \label{tbl-4} }
    \tablehead{
\colhead{   Galaxy  } &  \colhead{  $\theta_{ED}$   } &  \colhead{  $\Delta \alpha$ } &  \colhead{  $\Delta \delta$ } &  \colhead{  \vonez /\vone           } &  \colhead{  \vtwo /\vone            } &  \colhead{  \vthree /\vone          } \\
\colhead{       } &  \colhead{  (\arcsec )  } &
\multicolumn{2}{c}{    (\arcsec )          } &  \colhead{
} &  \colhead{              } &  \colhead{              }
    }
\startdata
    {\it Non-thermal} Model\tablenotemark{b}    &       &       &       &   0.46            &   0.56            &   0.18            \\
    {\it Thermal} Model\tablenotemark{c}    &       &       &       &   0.21            &   0.082           &   0.0057          \\
\\
    M82 &   $5.5 \times 5.6$\tablenotemark{d}   &   31.6    &   9.6 & $ 0.21    \pm 0.12    $ & $   0.40    \pm 0.13    $ & \nodata         \\
        &       &   26.3    &   8.0 & $ 0.23    \pm 0.07    $ & $   0.20    \pm 0.04    $ & \nodata         \\
        &       &   21.0    &   6.4 & $ 0.24    \pm 0.06    $ & $   0.26    \pm 0.06    $ & \nodata         \\
        &       &   15.8    &   4.8 & $ 0.35    \pm 0.05    $ & $   0.29    \pm 0.04    $ & \nodata         \\
        &       &   10.5    &   3.2 & $ 0.33    \pm 0.05    $ & $   0.32    \pm 0.05    $ & \nodata         \\
        &       &   5.3 &   1.6 & $ 0.44    \pm 0.13    $ & $   0.33    \pm 0.06    $ & \nodata         \\
        &       &   0.0 &   0.0 & $ 0.38    \pm 0.21    $ & $   0.47    \pm 0.10    $ & \nodata         \\
        &       &   -5.3    &   -1.6    & $ 0.39    \pm 0.14    $ & $   0.23    \pm 0.06    $ & \nodata         \\
        &       &   -10.5   &   -3.2    & $ 0.38    \pm 0.11    $ & $   0.33    \pm 0.05    $ & \nodata         \\
        &       &   -15.8   &   -4.8    & $ 0.47    \pm 0.06    $ & $   0.30    \pm 0.03    $ & \nodata         \\
        &       &   -21.0   &   -6.4    & $ 0.45    \pm 0.06    $ & $   0.27    \pm 0.04    $ & \nodata         \\
        &       &   -26.3   &   -8.0    & $ 0.65    \pm 0.20    $ & $   0.29    \pm 0.11    $ & \nodata         \\
        &       &   -31.6   &   -9.6    & $ 0.35    \pm 0.15    $ & $   0.21    \pm 0.09    $ & \nodata         \\
        &   11.1    &   14  &   6   &   \nodata         & $ 0.62    \pm 0.24    $ & \nodata         \\
        &       &   0   &   0   &   \nodata         & $ 0.75    \pm 0.15    $ & \nodata         \\
\\
    NGC 253 &   11.1    &   0   &   0   &   \nodata         & $ 0.34    \pm 0.15    $ & \nodata         \\
        &   81  &   0   &   0   &   \nodata         & $ 0.61    \pm 0.27    $ & $   0.18    \pm 0.09    $ \\
\\
    NGC 6946    &   11.1    &   0   &   0   &   \nodata         & $ 0.30    \pm 0.17    $ & \nodata         \\
\enddata
    \tablenotetext{a}{ We correct the extinction difference for each line, assuming the foreground extinction model (see Table~\ref{tbl-5}).}
    \tablenotetext{b}{ Fluorescent \hh\ emission for the PDR model 14 in \citet{Black87}. }
    \tablenotetext{c}{ Thermally excited \hh\ emission in LTE with $T_{exc} = 2000$ K \citep{Shull82,Black87}. }
    \tablenotetext{d}{ Size of the rectangular box of 5\farcs5 (major axis) $\times$ 5\farcs6 (minor axis). }
\end{deluxetable}

\begin{deluxetable}{lcccc}
    \tablewidth{0pt}
    \tabletypesize{\footnotesize}
    \tablecaption{ Total \hh\ \vone\ Flux and Extinction Correction  \label{tbl-5} }
    \tablehead{
\colhead{   Galaxy  } &  \colhead{  Obs. Flux\tablenotemark{a}  } &  \colhead{  \av\tablenotemark{b}    } &  \colhead{  ref } &  \colhead{  Corr. Flux  } \\
\colhead{       } &  \colhead{  ($10^{-17}$ W m$^{-2}$) } &
\colhead{  (mag)   } &  \colhead{      } &  \colhead{  ($10^{-17}$
W m$^{-2}$) }
    }
\startdata
    M82 &   49.5    &   5.5\tablenotemark{c}    &   1,2 &   86.9    \\
        &       &   52\tablenotemark{d} &   3   &   9820    \\
    NGC 253 &   31.7    &   7.8\tablenotemark{c}    &   4   &   69.9    \\
    NGC 6946    &   2.42    &   4.3\tablenotemark{c}    &   5   &   3.74    \\
    IC 342  &   1.71    &   10\tablenotemark{c,e}   &   6   &   3.68    \\
\enddata
    \tablenotetext{a}{Total \hh\ \vone\ flux from the observed regions in the galaxy.}
    \tablenotetext{b}{V-band extinction in magnitudes.}
    \tablenotetext{c}{Based on the foreground extinction model.}
    \tablenotetext{d}{Based on the mixed extinction model.}
    \tablenotetext{e}{Average value from the four regions in \citet{Boeker97}.}
\tablerefs{
    (1) \citealt{Lester90};
    (2) \citealt{Puxley91};
    (3) \citealt{Foerster01};
    (4) \citealt{Engelbracht98};
    (5) \citealt{Engelbracht96};
    (6) \citealt{Boeker97}
}
\end{deluxetable}

\begin{deluxetable}{lclclclcl}
    \tablewidth{0pt}
    \tabletypesize{\scriptsize}
    \tablecaption{ Observed Flux of Far-IR, \oi, and \cii  \label{tbl-6} }
    \tablehead{
\colhead{   Galaxy  }& \colhead{    $F_{H2}$    }& \colhead{    ref }& \colhead{    $F_{FIR}$\tablenotemark{a}  }& \colhead{    ref }& \colhead{    $F_{CII}$   }& \colhead{    ref }& \colhead{    $F_{OI}$    }& \colhead{    ref } \\
\colhead{       }& \multicolumn{2}{c}{  ($10^{-17}$ \wattmeter )            }& \multicolumn{2}{c}{  ($10^{-12}$ \wattmeter )            }& \multicolumn{2}{c}{  ($10^{-15}$ \wattmeter )            }& \multicolumn{2}{c}{  ($10^{-15}$ \wattmeter )            } \\
    }
\startdata
    M82 &   86.9    &   1   &   82.1    &   2,3,4,5,6,7 &   128 &   4   &   169 &   4   \\
    NGC 253 &   69.9    &   1   &   64.7    &   4,5,6,7 &   51.9    &   4   &   37.6    &   4   \\
    NGC 6946    &   3.74    &   1   &   3.68    &   3,4,5,6,7   &   10.3    &   4   &   5.90    &   4   \\
    IC 342  &   3.68    &   1   &   12.2    &   2,3,5,7 &   38.7    &   8   &   32.6    &   8   \\
\enddata
    \tablenotetext{a}{Average over the published values.}
\tablerefs{
    (1) This work;
    (2) \citealt{Crawford85};
    (3) \citealt{Stacey91};
    (4) \citealt{Negishi01};
    (5) \citealt{Young89};
    (6) \citealt{Smith96};
    (7) \citealt{Mouri92};
    (8) \citealt{Eckart90}
    }
\end{deluxetable}

\begin{deluxetable}{lccccccc}
    \tablewidth{0pt}
    \tabletypesize{\scriptsize}
    \tablecaption{Results from PDR Models \label{tbl-7} }
    \tablehead{
\colhead{   Galaxy  } &  \colhead{  log $n_H$ \tablenotemark{a} } &  \colhead{  log $G_\circ$ \tablenotemark{a} } &  \colhead{  log $I_{H2} / I_{FIR}$ \tablenotemark{b}    } \\
\colhead{       } &  \colhead{  (cm$^{-3}$) } &  \colhead{      }
&  \colhead{      }
    }
\startdata
    M82 &   2.7 &   2.9 & $ -5.2    $ \\
    NGC 253 &   1.8 &   2.6 & $ -5.6    $ \\
    NGC 6946    &   2.5 &   2.3 & $ -4.9    $ \\
    IC 342  &   2.8 &   2.5 & $ -4.8    $ \\
\enddata
    \tablenotetext{a}{Deduced physical parameters by comparing to the model results in \citet{Kaufman99}.}
    \tablenotetext{b}{Ratio of \hh\ \vone\ to far-IR intensity. $I_{H2}$ is derived from the PDR model in section~\ref{sec:model} and $I_{FIR}$ is from equation~(\ref{eq:G0_to_IFIR}).  }
\end{deluxetable}

\begin{deluxetable}{lccccc}
    \tablewidth{0pt}
    \tabletypesize{\footnotesize}
    \tablecaption{ Portion of a \hh\ Line in the Total \hh\ Emission  \label{tbl-8} }
    \tablehead{
\colhead{   Model   } &  \colhead{  Symbol  } &  \colhead{  \vone   } &  \colhead{  \vonez  } &  \colhead{  \vtwo   } &  \colhead{  \vthree }  \\
\colhead{       } &  \colhead{      } &  \colhead{  2.12183
\micron } &  \colhead{  2.22329 \micron } &  \colhead{  2.24771
\micron } &  \colhead{  2.20139 \micron }
    }
\startdata
    Non-thermal \tablenotemark{a}   &   $p_v$\tablenotemark{c}  &   0.016   &   0.0073  &   0.0089  &   0.0028  \\
    Thermal \tablenotemark{b}   &   $s_v$\tablenotemark{d}  &   0.085   &   0.018   &   0.0070  &   0.00048     \\
\enddata
    \tablenotetext{a}{ From the PDR model 14 in \citet{Black87}.  }
    \tablenotetext{b}{ From the LTE model with $T_{exc} = 2000$ K \citep{Black87}.  }
    \tablenotetext{c}{ $p_v = I_v^{nth} / I_T^{nth}$ }
    \tablenotetext{d}{ $s_v = I_v^{th} / I_T^{th}$ }
\end{deluxetable}


\end{document}